\documentclass[twocolumn,aps,prl,showpacs,amssymb,amsmath,superscriptaddress,floatfix]{revtex4}
\usepackage{graphicx}
\usepackage{epstopdf}
\usepackage{times}

\begin{document}

\title{Nearly Massless Electrons in the Silicon Interface with a Metal Film}

\author{Keun Su Kim}\affiliation{Center for Atomic Wires and Layers, Pohang University of Science and
Technology, Pohang 790-784, Korea}

\author{Sung Chul Jung}\affiliation{Department of Physics, Pohang University of Science and
Technology, Pohang 790-784, Korea}

\author{Myung Ho Kang}\affiliation{Department of Physics, Pohang University of Science and
Technology, Pohang 790-784, Korea}

\author{Han Woong Yeom}\email[]{yeom@postech.ac.kr}\affiliation{Center for Atomic Wires and Layers, Pohang University of Science and
Technology, Pohang 790-784, Korea}\affiliation{Department of
Physics, Pohang University of Science and Technology, Pohang
790-784, Korea}

\date{\today}

\begin{abstract}

We demonstrate the realization of nearly massless electrons in the
most widely used device material, silicon, at the interface with a
metal film. Using angle-resolved photoemission, we found that the
surface band of a monolayer lead film drives a hole band of the Si
inversion layer formed at the interface with the film to have nearly
linear dispersion with an effective mass about 20 times lighter than
bulk Si and comparable to graphene. The reduction of mass can be
accounted for by repulsive interaction between neighboring bands of
the metal film and Si substrate. Our result suggests a promising way
to take advantage of massless carriers in silicon-based thin-film
devices, which can also be applied for various other semiconductor
devices.

\end{abstract}

\pacs{}
\maketitle
\newpage


The ultimate performance of electronic devices is largely governed
by the effective mass of charge carriers. Therefore, the
never-ending quest for higher performance devices has looked after
high speed carriers with light effective mass \cite{SERVICE}. This
quest may have reached its ultimate goal through the recent findings
of massless electrons with their speed close to light in graphene
\cite{ELI,NOV} and bismuth compounds \cite{LI,BISB,BISE,BITE}.
However, for these materials various difficulties exist in promptly
realizing practical and mass-producible devices.


In general, the effective mass of electrons is determined by their
energy dispersion. Especially, the edge structures of conduction and
valence bands near the energy gap ($E_{g}$) are important. In the
band-edge region, the dispersion is sensitive to the interaction
between the bottom conduction and the top valence band as described
well by the $k\cdot{p}$ theory \cite{DAVIES}. This standard theory
provides an approximate analytic expression for band dispersion
under the interband interaction around high-symmetry points of
momentum ($k$) space \cite{DAVIES}. The interaction energy,
$\pm$$\sqrt{(E_{g}/2)^{2}+(P\cdot{k})^{2}}$ ($P$, the optical matrix
element), gives rise to repulsion between bands and provides the
linear component in dispersion \cite{DAVIES}. As $E_{g}$ approaches
zero, the dispersion becomes nearly linear $E$($k$) $\approx$
$P\cdot{k}$ with a much lighter effective mass of electrons
$m^{\ast}$ = $\hbar^{2}$$E_{g}/(2P^{2})$. This effect explains the
trend that a semiconductor with a narrower gap has a lighter
effective mass \cite{YU} and the zero-gap semiconductor of graphene
has a negligible effective mass \cite{ELI,NOV}. This theory also
tells us that if the energy gap could be controlled, the effective
mass would be tuned through the interband interaction. While the
energy gap of bulk materials cannot be tuned easily as fixed by
their crystalline structure, here we show that a similar effect can
be obtained at a semiconductor interface where a proper interface
state is formed within the band gap \cite{CHIANG,CHIANG2,MORAS}.

\begin{figure}
\includegraphics[scale=1]{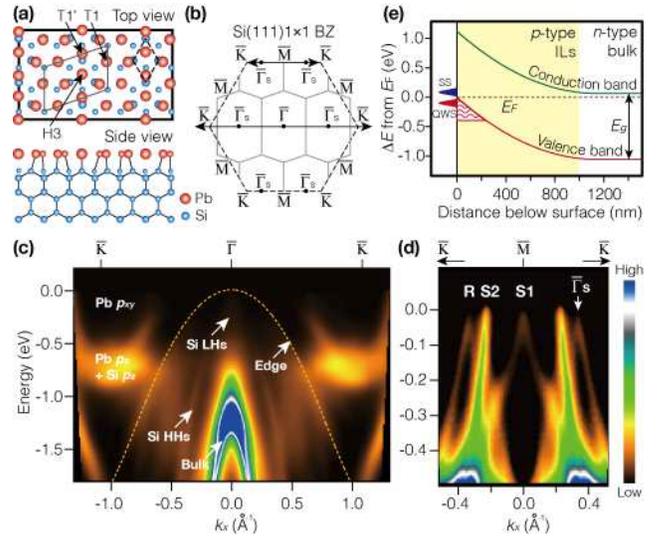}
\caption{\label{Fig1}(color online). (a) Crystal- and (b)
reciprocal- lattice structures for Pb/Si(111) at 1.2 ML with a
$\sqrt{7}\times\sqrt{3}$ symmetry \cite{BROCHARD}. Grey (dashed)
line represents the $\sqrt{7}\times\sqrt{3}$ (1$\times$1) surface
unit cell. ARPES data collected along (c) long and (d) short arrows
in (b) crossing 1$\times$1 ($\bar{\Gamma}$) and
$\sqrt{7}\times\sqrt{3}$ ($\bar{\Gamma}_{S}$) Brillouin zone
centers, respectively. The dashed line in (c) denotes the Si
valence-band edge. $\Lambda$-shaped band with a linear dispersion on
each $\bar{\Gamma}_{S}$ in (d) is labeled R and two other gap-state
bands in between are denoted by S1 and S2. Both ARPES images are
symmetrized with respect to the origin. (e) Band bending in the
present $n$-type Si substrate calculated by the measured $E_{F}$
position and the bulk doping concentration \cite{MONCH}.}
\end{figure}

We chose an ultrathin (only single layer) Pb film on $n$-type
Si(111) substrate, which has ideally abrupt interfacial structure
and strongly dispersing metallic electron bands within the Si band
gap \cite{BROCHARD,CUDAZZO,CHOI}. For this system, the
film-substrate interaction, which is exploited in the present work,
was already invoked to explain the anomalous superconductivity at
the two-dimensional (2D) limit \cite{QKXUE,SHIH}. Figure
\ref{Fig1}(a) illustrates the atomic structure of the interface with
the Pb density of 1.2 monolayer (ML) (7.84 atoms/nm$^{2}$). Pb atoms
are densely packed within a single layer upon the bulk-terminated
Si(111) surface \cite{BROCHARD}. Most of Pb atoms sit on top of
underlying Si atoms (T1 or T1$^{\prime}$ sites) while part of them
are slightly displaced (T1$^{\prime}$) by additional Pb atoms in
hollow sites (H3), which lead to the formation of a uniform
$\sqrt{7}\times\sqrt{3}$ unit cell (grey lines) \cite{SUPP}. Due to
the anisotropy of this unit cell and the three-fold symmetry of the
substrate, one inevitably obtains triply rotated domains
\cite{CHOI,SUPP}, which should be considered in interpreting
experimental data.

Angle-resolved photoemission (ARPES) measurements were conducted in
an ultra-high vacuum chamber (6.5 $\times$ 10$^{11}$ torr) equipped
with a hemispherical electron analyzer (R4000, VG Scienta) and a
high-flux He discharge lamp for 21.2-eV photons. The samples were
cryogenically cooled down to 90--100 K for measurements. The overall
energy and momentum resolutions were better than 20 meV and 0.02
{\AA}$^{-1}$. Figure \ref{Fig1}(c) shows band dispersions along the
high-symmetry direction through the center of Brillouin zone
$\bar{\Gamma}$ [the long arrow in Fig. \ref{Fig1}(b)] measured by
ARPES. Near $\bar{\Gamma}$, an intense parabolic band is readily
found, which is the well-known direct transition from bulk Si
valence bands \cite{UHRBERG}. Another feature with strong intensity
appears near each zone boundary ($\bar{K}$) with little dispersion
around 0.7 eV, which is the covalent-type bonding state between Pb
and Si \cite{CUDAZZO}. In contrast, around $\bar{K}$, there are
parabolic bands dispersing toward the Fermi energy ($E_{F}$), which
were identified as due to 2D metallic electrons localized within the
Pb layer (called S2 hereafter) \cite{CHOI}. This metallic surface
state induces a huge upward band bending in the $n$-type substrate,
and forms a $p$-type inversion layer (IL) as shown in Fig.
\ref{Fig1}(e) \cite{MONCH}. The strong band bending yields a
triangular potential well to confine electrons in the interface IL
into 2D quantum well states (QWS) \cite{CHIANG,MONCH}. An IL state
at interfacial layers is typically not easily probed by very surface
sensitive ARPES but, for an ultrathin metal film like single-layer
In on Si(111), a previous ARPES study traced its detailed dispersion
\cite{TAKEDA}. We also found such IL states; two hole-like bands are
observed with weak intensity between the bulk and surface bands in
Fig. \ref{Fig1}(c). A relatively prominent band is identified as the
so called heavy-hole (HH) bands \cite{TAKEDA} but the other branch
with a lighter effective mass, light-hole (LH) band, is barely
observable as a weak and broad feature on top of the Si bulk band.

\begin{figure}
\includegraphics[scale=1]{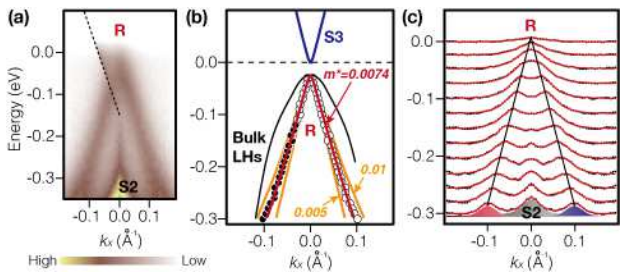}
\caption{(color online). (a) ARPES data of the R band. Data below
the dashed line are those symmetrized with respect to
$\bar{\Gamma}_{S}$ to eliminate the strong neighboring feature S2.
(b) Spectral peak positions of the R band (open circles) extracted
from MDCs shown in (c). Data points given in closed circles are
those symmetrized. The dispersion of a normal Si LH band (black
parabola) is compared to show the reduction of curvature. Red and
blue lines are the results of $k\cdot{p}$ model fit \cite{EQ}. The
error bar in $k$ is less than $\pm$0.02 {\AA}$^{-1}$, which brackets
the quantified effective mass (orange lines) (see also the
supplementary information \cite{SUPP}). Red lines overlaid in (c)
show fits by Lorenzian functions.}\label{Fig2}
\end{figure}


Although the LH band is not clear enough here [Fig. \ref{Fig1}(c)]
due to its weak intensity, we can map its dispersion clearly at the
center of the surface Brillioun zone ($\bar{\Gamma}_{S}$) away from
$\bar{\Gamma}$ as translated by the surface periodicity
\cite{MORAS,TAKEDA}. Figure \ref{Fig1}(d) shows band dispersions
taken through two such $\bar{\Gamma}_{S}$'s and the $\bar{M}$ point
[the short arrow in Fig. \ref{Fig1}(b)]. There are strongly
dispersing bands with dominant intensity, crossing $E_{F}$ at
$\pm$0.21 {\AA}$^{-1}$, which correspond to the S2 state for Pb
metallic electrons. In addition, two other bands are identified, S1
folded back near $E_{F}$ with respect to $\bar{M}$ and R folded with
respect to $\bar{\Gamma}_{S}$. Surprisingly, their dispersions are
$\Lambda$-shaped and apparently very linear, especially for R. The
detailed dispersion of R, located on $\bar{\Gamma}_{S}$ and related
to the Si LH band below, can be shown more clearly and quantified by
the peak positions of momentum distribution curves (MDCs) in Fig.
\ref{Fig2}(c). The dispersion is indeed linear within the
experimental uncertainty as shown in Fig. \ref{Fig2}(b) (open
circles). Thus, the effective mass value would be extremely small
and cannot be quantified by a simple parabolic fit of the dispersion
as in the case of graphene \cite{ELI} and Bi compounds
\cite{BISB,BISE,BITE}. Instead, the Fermi velocity can
straightforwardly be extracted from the slope of the band and is as
high as 4.6 $\pm$ 0.4 $\times$ 10$^{5}$ m/s, which reaches to half
of those in graphene \cite{ELI,SUPP} and Bi$_{0.9}$Sb$_{0.1}$
\cite{BISB}, and is similar to those in Bi$_{2}$Se$_{3}$ \cite{BISE}
and Bi$_{2}$Te$_{3}$ \cite{BITE}. The effective mass value of this
linear band will be discussed further below.

\begin{figure}
\includegraphics[scale=1]{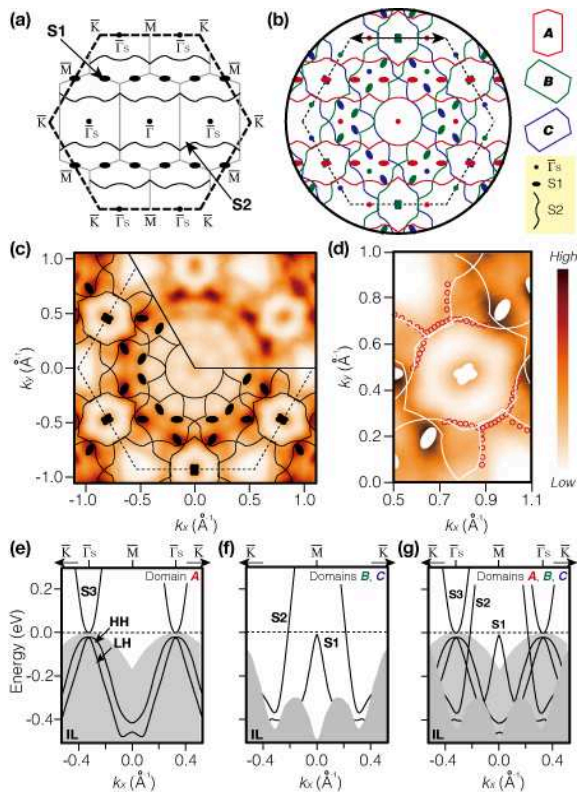}
\caption{(color online). Constant energy contours at $E_{F}$
calculated from DFT for the (a) single-domain and (b) triply-domain
surfaces. Contours from each domain ($A$, $B$, and $C$) are
indicated by different colors in (b). (c) ARPES constant energy
contours at $E_{F}$. The DFT contours from (b) are superimposed. The
raw data (region without DFT contours) are symmetrized reflecting
the fundamental mirror symmetry. (d) Enlarged data around $\bar{M}$
for the detailed comparison between theory (while lines) and
experiment (red circles extracted by MDC peak positions). DFT band
dispersions along the arrow in (b) for (e) domain $A$ and (f)
domains $B$ and $C$, which are overlapped together in (g). The
shaded area represents the projected bulk band. For the best match
with the experimental data the calculated surface state energies are
rigidly shifted by 120 meV with respect to Si bands.}\label{Fig3}
\end{figure}


Since the R band with a striking dispersion cannot be found in Si
bulk crystals [Fig. \ref{Fig2}(b)] and on clean Si(111) surfaces
\cite{UHRBERG}, it must be due to electron states localized at the
interface (Si ILs) or within the Pb layer. To identify the origin of
each band further, we performed theoretical calculations based on
density functional theory (DFT). The calculations were performed
using the Vienna \textit{ab-initio} simulation package within the
generalized gradient approximation and the ultrasoft pseudopotential
scheme with a plane wave basis \cite{KRESSE}. The surface is modeled
by a periodic slab geometry with 6 and 12 Si layers whose bottoms
are terminated by hydrogen.


Figure \ref{Fig3}(a) shows the calculated constant energy contours
at $E_{F}$. There are two kinds of surface-state Fermi contours,
point-like crossings and wavy lines repeated following the surface
periodicity. Since ARPES detects signals coming from all
triply-rotated domains, the expected Fermi contours result in a
complex pattern [Fig. \ref{Fig3}(b)]. Nevertheless, the agreement
between the calculated and experimental Fermi contours is excellent
in Figs. \ref{Fig3}(c) and \ref{Fig3}(d). The point-like crossings
and the details of the wavy contours are all precisely reproduced,
which correspond to S1 and S2, respectively. The dispersions of S1
and S2 are also reproduced quantitatively well in the calculation
[Fig. \ref{Fig3}(f)]. These states come from the Pb layer due to
in-plane Pb 5$p$ orbitals. In fact, S1 and S2 bands are
doubly-degenerated with contributions from domains $B$ and $C$
overlapped exactly [Fig. \ref{Fig3}(f) and green and blue ones in
Fig. \ref{Fig3}(b)]. The contribution from domain $A$, which pass
through two $\bar{\Gamma}_{S}$ points [Fig. \ref{Fig3}(e) and red
ones in Fig. \ref{Fig3}(b)], is not degenerated. For this domain,
the Pb-derived band (S3) is located just above $E_{F}$ and two
Si-derived bands below $E_{F}$ at $\bar{\Gamma}_{S}$ [Fig.
\ref{Fig3}(e)]. The calculated electron density for the latter is
distributed within Si layers and their dispersions are consistent
with the Si IL states (LH and HH) mentioned above. In particular,
one with a sharper dispersion (LH) is very similar to R as compared
in Fig. \ref{Fig2}(b). The calculation, thus, clearly identifies S1,
S2, S3 and R bands as three 5$p$ states of the Pb layer and a
Si-derived state, and reproduces the dispersions of S1, S2, and S3
quantitatively well. This result is consistent with the very recent
work in Ref. \cite{HSU} for the single-domain surface.


One notable feature of the calculation is that the bottom of the S3
band lies on $E_{F}$, namely, just above the experimentally measured
band R with a very small energy separation. Thus, S3 has a chance to
interact strongly with R from Si subsurface layers through the
interband interaction \cite{DAVIES}. As mentioned above, this
interband interaction determines the effective mass of the bands
according to the $k\cdot{p}$ model \cite{DAVIES}. Indeed, such an
interaction between the band of a Ag monolayer film and the band of
a Si(111) substrate was observed recently and affects the dispersion
of a S3-like surface state of Ag \cite{LIU}.

The fitting of our experimental dispersion using the $k\cdot{p}$
model \cite{EQ} with the band-gap size determined by the calculated
energy position of S3, 23 meV, reproduces well the linear dispersion
of the R band as shown in Fig. \ref{Fig2}(b), yielding an extremely
light effective mass of 0.0074 $\pm$ 0.0015$m_{e}$, where $m_{e}$ is
the electron rest mass. In fact, the $k\cdot{p}$ fitting is simple
enough to treat the band-gap size as another free parameter and the
band gap is reliably determined to be within 18--24 meV \cite{SUPP}
in good accord to the theoretical value. The effective mass
determined is 20 times lighter than the normal Si LH band
(0.15$m_{e}$) \cite{YU}. Note also that, for the HH bands, the
interband interaction is forbidden from the band symmetry
\cite{DAVIES}. Therefore, we conclude that the R band with an
extremely linear dispersion is the modified LH band through the
interband interaction with S3, while we do not know why the HH band
itself is not observed around $\bar{\Gamma}_{S}$.


The $k\cdot{p}$ method has been widely applied to various
semiconductors and the smallest effective mass reported is
0.016$m_{e}$ for InSb \cite{YU,SUPP}. This yields the carrier
mobility as high as 30,000 cm$^{2}$ V$^{-1}$ s$^{-1}$, about 2
orders of magnitude higher than bulk Si \cite{SERVICE}. The present
effective mass for the Si IL, thus, indicates the possibility of an
ultrafast carriers at Si interfaces. This method has not been
applied to graphene and Bi compounds, for which only the effective
mass values measured by transport experiments are available; about
(0.002--0.009)$m_{e}$ in a similar range with the present value
\cite{NOV,BISB,SUPP}. In contrast, the DFT calculation shown above
does not reproduce exactly the linear dispersion of the LH band,
yielding an effective mass of roughly 0.07$m_{e}$. We think this
discrepancy is due to the limitation of the present slab calculation
with only few Si layers in reproducing properly the QWS of the much
thicker IL.

\begin{figure}
\includegraphics[scale=0.95]{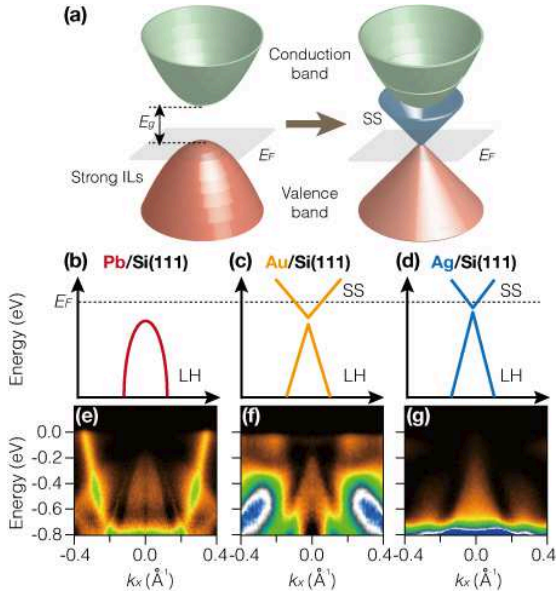}
\caption{(color online). (a) Schematic illustration of the mechanism
for a linear hole band dispersion in semiconductor ILs. (b)--(d)
Schematics of surface (SS) and LH band dispersions near
$\bar{\Gamma}$ in Pb, Au, and Ag single-layer films on Si(111) with
a common $\sqrt{3}\times\sqrt{3}$ symmetry \cite{CHOI,LIU,HIMPSEL}.
(e)--(g) Corresponding ARPES dispersions of LH bands measured along
the long arrow in Fig. \ref{Fig1}(b) across
$\bar{\Gamma}$.}\label{Fig4}
\end{figure}

The proposed mechanism for the linear dispersion is summarized in
Fig. \ref{Fig4}(a). The contact of a metal film to a semiconductor
substrate plays two key roles; (i) to induce a strong IL near the
interface lifting the hole bands of the substrate up close to
$E_{F}$ and (ii) to provide a proper band (S3 here) within the band
gap inducing the strong repulsive interaction with hole bands for
linear dispersion. In order to verify this mechanism, we took
advantage of diverse surface band structures in metal/Si(111)
systems. We examined three representative metal overlayers --- Pb,
Au, and Ag on Si(111) --- with a common $\sqrt{3}\times\sqrt{3}$
symmetry but with different surface bands. Figures
\ref{Fig4}(b)--\ref{Fig4}(d) describe corresponding band dispersions
around $\bar{\Gamma}$ taken from previous reports in literature
\cite{CHOI,LIU,HIMPSEL}. The $\sqrt{3}\times\sqrt{3}$-Pb system
(with the Pb density of 4/3 ML) has no bands close to Si LH bands
\cite{CHOI} while the others (at 1 ML) have adequate bands from the
films similar to the present case to allow the interaction with the
LH band (band gaps of 40--60 meV) \cite{HIMPSEL,JEONG}. The
consequence of such a difference on the dispersion of the LH band is
remarkable. In Fig. \ref{Fig4}(e), our ARPES data for the Pb system
show a normal parabolic dispersion consistent with bulk Si
\cite{DAVIES}. However, for Au and Ag systems, dispersions are
obviously much more linear [Figs. \ref{Fig4}(f) and \ref{Fig4}(g)].
These results not only provide solid evidence on the validity of the
proposed mechanism but also suggest the possibility of applications
for a wider range of materials.

Our result suggests an unprecedented way to ultrafast electronic
devices based on Si. Since this technique directly controls the
carrier mass at the interface by a metal overlayer, it does not
require any artificial modification or chemical engineering on the
Si bulk lattice itself. Furthermore, this mechanism, in principle,
can be generally applied to various other semiconductors even with
thicker films \cite{CHIANG2}, provided that proper interface states
exist. For example, this may solve the notorious imbalance between
the hole and electron mobility of III-V compounds \cite{SERVICE}.

This work was supported by KRF through the CRi program. M.H.K.
acknowledges the support from KRF (Grant No. 2009-0074825).



\begin{references}
\bibitem{EQ} The dispersions of upper and lower bands are given by $E$($k$) = $E_{0}$ $\mp$ ${E_{g}}/2$ + $\frac{\hbar^{2}k^{2}}{2m_{e}}$ $\pm$ $\sqrt{(E_{g}/2)^{2}+(P\cdot{k})^{2}}$.
\end{references}
\end{document}